\documentclass[fleqn,10pt]{wlscirep}
\title{PsiQuaSP -- A library for efficient computation of symmetric open quantum systems}

\usepackage{amssymb, amsmath}
\usepackage{mathtools}
\usepackage[english]{babel}
\usepackage{bm}
\usepackage{graphicx,accents}
\usepackage{epstopdf}
\usepackage{lmodern}
\usepackage[utf8]{inputenc}
\usepackage{color}
\usepackage{comment}
\usepackage{stmaryrd}
\usepackage[super]{natbib}

\makeatletter
\DeclareRobustCommand{\cev}[1]{%
  \mathpalette\do@cev{#1}%
}
\newcommand{\do@cev}[2]{%
  \fix@cev{#1}{+}%
  \reflectbox{$\m@th#1\vec{\reflectbox{$\fix@cev{#1}{-}\m@th#1#2\fix@cev{#1}{+}$}}$}%
  \fix@cev{#1}{-}%
}
\newcommand{\fix@cev}[2]{%
  \ifx#1\displaystyle
    \mkern#23mu
  \else
    \ifx#1\textstyle
      \mkern#23mu
    \else
      \ifx#1\scriptstyle
        \mkern#22mu
      \else
        \mkern#22mu
      \fi
    \fi
  \fi
}

\makeatother

\author[1,*]{Michael Gegg}
\author[1]{Marten Richter}
\affil[1]{Institut für Theoretische Physik, Nichtlineare Optik und Quantenelektronik, Technische Universität Berlin, Hardenbergstr. 36, EW 7-1, 10623 Berlin, Germany}
\affil[*]{michael.gegg@tu-berlin.de}

\begin{abstract}
In a recent publication we showed that permutation symmetry reduces the numerical complexity of Lindblad quantum master equations for identical multi-level systems from exponential to polynomial scaling. This is important for open system dynamics including realistic system bath interactions and dephasing in, for instance, the Dicke model, multi-$\Lambda$ system setups etc. Here we present an object-oriented C++ library that allows to setup and solve arbitrary quantum optical Lindblad master equations, especially those that are permutationally symmetric in the multi-level systems. PsiQuaSP (Permutation symmetry for identical Quantum Systems Package) uses the PETSc package for sparse linear algebra methods and differential equations as basis. The aim of PsiQuaSP is to provide flexible, storage efficient and scalable code while being as user friendly as possible. It is easily applied to many quantum optical or quantum information systems with more than one multi-level system. We first review the basics of the permutation symmetry for multi-level systems in quantum master equations. The application of PsiQuaSP to quantum dynamical problems is illustrated with several typical, simple examples of open quantum optical systems.
\end{abstract}

\begin{document}

\flushbottom
%TC:ignore
\maketitle
%TC:endignore

\thispagestyle{empty}

\section{Introduction}
In quantum optics and more recently also quantum information research is often centered around how multiple quantum emitters or multi-level systems interact with each other and/or the (photonic) environment. Generally using an open system description is desirable, since dissipation and dephasing are omnipresent. In these systems many body effects produce a rich variety of physical effects but the full quantum description of many emitters usually results in an exponential complexity for numerical treatments. Since exact analytic solutions in such systems are rare and a straightforward numerical treatment of systems of exponential complexity is limited  to very small systems alternative methods are necessary.\\
In a recent publication we have shown that identical emitters in quantum optical Lindblad master equations result in a permutation symmetry that can be used to reduce the complexity from exponential to polynomial in the number of multi-level systems $N$\cite{Gegg:NJP:16}. The approach is exact/non-approximate and non-perturbative, which implies that the method is valid for any permutation symmetric master equation and all parameter ranges. For two-level systems the method has been used by various authors\cite{Sarkar:JPhysA:87,Sarkar:EuroPhysLett:87,Carmichael::02,Hartmann:QIC:16,Xu:PhysRevA:13,Richter:PhysRevB:15,Kirton:PhysRevLett:17,Gegg:arxiv:17}. Examples for compatible open system setups that can be described are Dicke super- and subradiance\cite{Dicke:PhysRev:54,Garraway:ptrsl:11,Gegg:arxiv:17}, lasers and related devices\cite{Richter:PhysRevB:15}, theoretical toy models such as the open Lipkin-Meshkov-Glick model\cite{Lipkin:NuclPhys:65,Lee:PhysRevA:14}, many particle contributions to STIRAP or coherent population trapping in three-level systems\cite{Bergmann:RevModPhys:98}, multi-biexciton cascades in quantum dots\cite{Hein:PhysRevLett:14} and others. The method allows to study all these systems, including realistic dephasing and dissipation while giving full access to the complete density matrix and thus all information about the system. The mentioned quantum systems are studied in many different contexts including different types of phase transitions\cite{Wang:PhysRevA:73,Emary:PhysRevLett:03,Walls:SPTP:78,Lee:PhysRevA:14,Gegg:arxiv:17}, generation of quantum light\cite{Hein:PhysRevLett:14,Kuhn:OE:16}, lasing\cite{Genway:PhysRevLett:14,Richter:PhysRevB:15}, entanglement\cite{Solano:PhysRevLett:03,Gonzales:PhysRevLett:13,Otten:PhysRevA:16}, super- and subradiance\cite{Scully:PhysRevLett:15,Gegg:arxiv:17} and quantum information storage\cite{Scully:PhysRevLett:15,Gegg:arxiv:17}. Our library allows to directly solve such master equations for moderate multi-level system numbers. This includes the study of quantum many body effects in the presence of dephasing, which was not feasible previously for these systems.\\
In this article we introduce a ready to use computer library for quantum optical master equations for systems of many identical emitters\cite{psiquasp}. The associated permutation symmetry allows to reduce the exponential complexity to polynomial without any approximation. The library is called \emph{PsiQuaSP -- Permutation symmetry for identical Quantum Systems Package}. The method fills the gap for few and intermediate multi-level system numbers $N$ left by quantum optical phase space methods such as the positive P representation\cite{Carmichael:PhysRevA:86,Sarkar:JPhysA:87,Sarkar:EuroPhysLett:87,Gilchrist:PhysRevA:97,Carmichael::02} that cover the large $N$, classical limit.\\
PsiQuaSP enables the setup of the master equation in computer code. The actual numerical solution is entirely handled by PETSc \cite{petsc-web-page,petsc-user-ref,petsc-efficient} and related packages such as SLEPc \cite{Hernandez:LNCS:03,Hernandez:TMS:05,slepc-users-manual}. These are state-of-the-art packages for efficient sparse linear algebra methods and differential equations. PETSc and SLEPc use MPI distributed memory parallelism. Additionally PETSc provides interfaces to many advanced, external libraries for e.g. specialized linear algebra tools and optimization of parallel performance like MUMPS\cite{mumps}, SuperLU\cite{Li:TMS:05}, METIS/ParMETIS\cite{parmetis}, PTScotch\cite{Chevalier:PC:08,ptscotch} and others. This ensures that PsiQuaSP users can always use current and most appropriate algorithms and can directly access the advanced computational sparse matrix methods available through PETSc. This makes PsiQuaSP scalable and versatile.\\
The paper is organized as follows: In section \ref{sec.ps} we give a quick introduction to the permutation symmetry methodology of Ref. \citenum{Gegg:NJP:16}. Especially we introduce sketches which facilitate the setup of the simulation. In section \ref{sec.basicst} we explain the basic design of the library, how it should be used and illustrate the application of the library using examples with two- and three-level systems. In section \ref{sec.overview} we give an overview over all ready-made Liouville operator templates in PsiQuaSP and explain how to construct custom types in section \ref{sec.sketches}. Finally in section \ref{sec.performance} we give a short discussion about the performance of the library.

\section{Lindblad master equations and permutation symmetry}
\label{sec.ps}

As stated in the previous section we target Lindblad master equations of collections of identical, indistinguishable multi-level systems. The prerequisite of identical, indistinguishable systems results in the permutation symmetry.\\
\emph{Notation:} We label the states of the individual multi-level system with integers starting from zero: $|0\rangle_i$, $|1\rangle_i$, $|2\rangle_i$, $\dots$. $|0\rangle_i$ is usually the ground state and the index $i$ refers to the individual system, which is sometimes just referred to as spin. We use general spin matrices describing the individual system/spin according to their Ket and Bra notation:
\begin{equation}
\sigma_{kl}^i = |k\rangle_i \langle l|_i.
\end{equation}
In contrast the direct product of $n$ spin matrices $\sigma_{kl}$, each referring to another multi-level system is denoted as
\begin{equation}
\sigma_{kl}^{\otimes n} = \underbrace{\sigma_{kl}^i \otimes \sigma_{kl}^j \otimes \dots}_{n ~\mbox{factors}}.
\end{equation}
The Liouville space basis for an individual two-level system is formed by four spin matrices, for three-level systems by nine matrices and for general $d$-level systems by $d^2$ spin matrices. General collective spin operators are defined as 
\begin{equation}
J_{kl} = \sum_i \sigma_{kl}^i,
\label{eq.defcollspinop}
\end{equation}
and bosonic operators (e.g. for photons, phonons etc.) are labeled as $b$ and $b^\dagger$. In the context of two-level systems it is customary to define the $\sigma_z^i = 1/2(\sigma_{11}^i-\sigma_{00}^i)$ and $J_{z}= \sum_i \sigma_z^i$ operators and also to use the labels $+,-$ instead of $10,01$. We do not use the two-level system notation in this report since it is confusing for multi-level systems and our aim is a clear and consistent notation for all types of multi-level systems.\\
\subsection{Examples for master equations}
\begin{figure}
\centering
\includegraphics[scale=0.8]{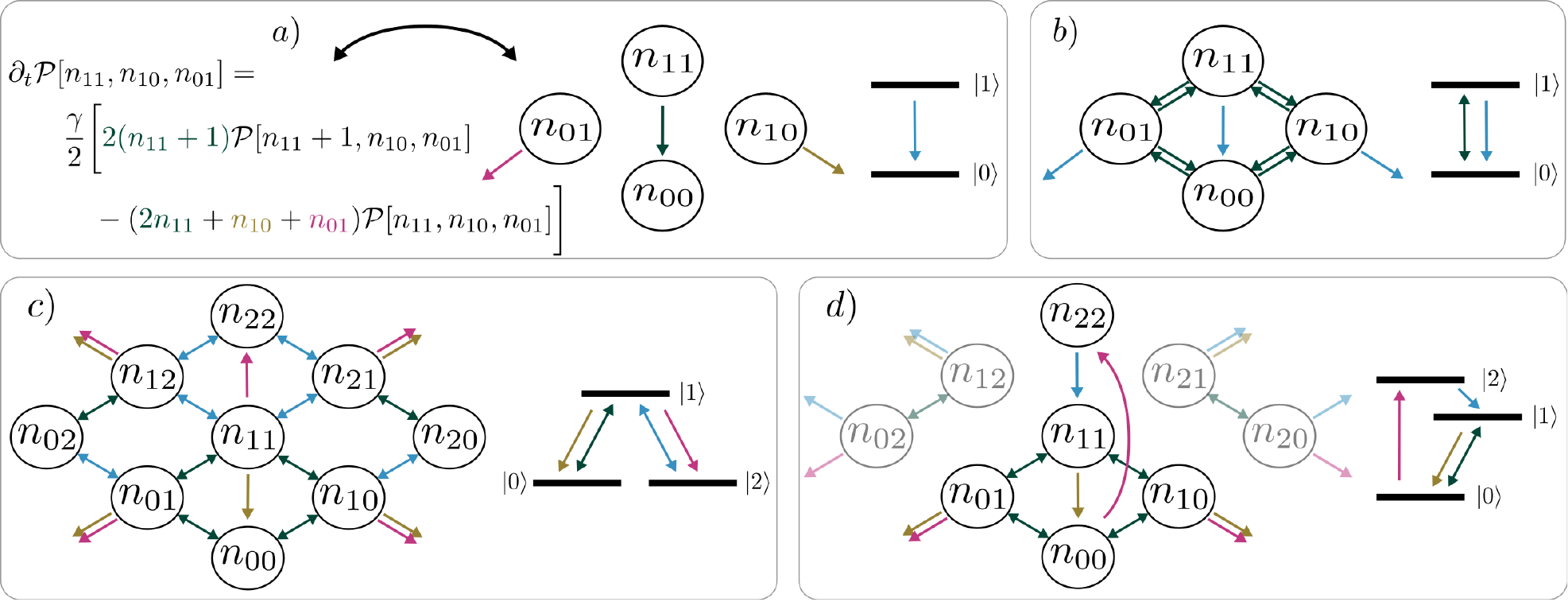}
\caption{Illustration of the processes of the master equations for two- and three-level systems (right side in a)-d) shows level schemes and left side shows corresponding sketches): a) Translating an equation into a sketch -- arrows and corresponding terms have the same color. The green arrow depicts the loss of excitation, states with $n_{11}+1$ decay into states with increased $n_{00}$ until reaching the ground state (i.e. $n_{11}=0,~n_{00}=N$). The yellow and purple arrows depict the dephasing. The offdiagonal elements ($n_{10},n_{01}\neq 0$) are just dephased. The arrows pointing to the outside indicate loss. b) Open Dicke/Tavis-Cummings model: Emitter-mode coupling part (green arrows) of equation \eqref{eq.dickeham} and individual spontaneous emission part (blue arrows) (equation \eqref{eq.tlsspontem}). c) $\Lambda$-system setup equations \eqref{eq.lambdaham}, \eqref{eq.lambdadiss}: Two different interactions from equation \eqref{eq.lambdaham} (green,blue) and two different spontaneous emission processes from equation \eqref{eq.lambdadiss} (yellow,purple). d) Three-level laser setup (Ref. \citenum{Gegg:NJP:16}): Population mechanism (pink,blue), coupling to the lasing mode (green) and spontaneous emission into nonlasing modes (yellow). Four coherence degrees of freedom ($n_{20}$, $n_{21}$, $n_{02}$ and $n_{12}$) are decoupled from the densities.}
\label{fig.mlsintro}
\end{figure}

A general Lindblad equation is defined as \cite{Breuer::02}
\begin{equation}
\partial_t \rho = \mathcal{L} \rho.
\end{equation}
One example for a master equation with permutational symmetry is the open Dicke model, i.e. a set of identical two-level systems coupled to a bosonic mode
\begin{equation}
\partial_t \rho = \frac{i}{\hbar}[\rho,H] + \mathcal{D}_1(\rho) + \mathcal{D}_2(\rho),
\label{eq.tlsmaster}
\end{equation}
with the self energy and interaction Hamiltonian (see Fig. \ref{fig.mlsintro} b) right)
\begin{equation}
H= \hbar \omega_0 b^\dagger b + \hbar \omega_{11} J_{11} + \hbar g (J_{01}+J_{10})(b^\dagger + b),
\label{eq.dickeham}
\end{equation}
and including e.g. the two-level system spontaneous emission and cavity lifetime Lindblad dissipator
\begin{align}
\mathcal{D}_1(\rho) & = \frac{\gamma}{2} \sum_i (2\sigma_{01}^i \rho \sigma_{10}^i - \sigma_{11}^i \rho - \rho \sigma_{11}^i), \nonumber\\
\mathcal{D}_2(\rho) &= \frac{\kappa}{2} (2b \rho b^\dagger - b^\dagger b \rho - \rho b^\dagger b).
\label{eq.tlsspontem}
\end{align}
The setup is permutationally symmetric since the two-level system parameters in this equation, i.e. $\omega_1$, $g$ and $\gamma$ are identical for all two-level systems. Exchanging the indices of any two two-level systems results in the same equation.\\
All quantum master equations for sets of multi-level systems, where the parameters in the master equation do not depend on the index of the individual multi-level systems show this permutation symmetry.
Another example would be a collection of $\Lambda$ systems, where for instance one transition is interacting with a bosonic mode and the other is driven by an external laser, see Fig. \ref{fig.mlsintro} c) right. In an appropriate rotating frame the Hamiltonian reads
\begin{equation}
H  =\hbar \Delta_0 b^\dagger b + \hbar \Delta_1 J_{22} + \hbar g (J_{01} b^\dagger+J_{10} b) + \hbar E (J_{21}+J_{12}),
\label{eq.lambdaham}
\end{equation}
where $\Delta_0$ is the detuning between the $0 - 1$ transition and the cavity mode and $\Delta_1$ is the detuning between the $1 - 2$ transition and the driving laser. Open system contributions are e.g. spontaneous emission and a finite photon lifetime
\begin{align}
\mathcal{D}_1(\rho) &= \frac{\gamma}{2} \sum_i (2\sigma_{01}^i \rho \sigma_{10}^i - \sigma_{11}^i \rho - \rho \sigma_{11}^i),\qquad
\mathcal{D}_2(\rho) = \frac{\gamma'}{2} \sum_i (2\sigma_{21}^i \rho \sigma_{12}^i - \sigma_{11}^i \rho - \rho \sigma_{11}^i),\nonumber\\
\mathcal{D}_3(\rho) &= \frac{\kappa}{2} (2 b \rho b^\dagger - b^\dagger b \rho - \rho b^\dagger b).
\label{eq.lambdadiss}
\end{align}
An analytic solution is only possible for very few of such equations and there are many different approximate schemes to attack this problem: phase space methods like positive P representation\cite{Gilchrist:PhysRevA:97,Carmichael::02,Orioli:arxiv:17}, limits like single excitation limit\cite{Feist:PRL:15} or reductions to the superradiant or general completely symmetric multiplet subspaces\cite{Hayn:PhysRevA:11,Lee:PhysRevA:14} and related techniques like Holstein-Primakoff transformation and -approximation\cite{Holstein:PhysRev:40,Garraway:ptrsl:11,Hayn:PhysRevA:11}, truncation of the hierarchy of operator expectation values -- also called cluster expansion\cite{Richter:JChemPhys:07,Schneebeli:PhysRevLett:08,Genway:PhysRevLett:14} or, more recently, matrix product state or matrix product operator based truncation schemes explicitly for spin-boson models\cite{Wall:PhysRevA:16}. There are also nonapproximate approaches like quantum trajectory/quantum-jump Monte Carlo approaches\cite{Dalibard:PhysRevLett:92,Carmichael::09}. All these methods have their advantages and drawbacks, together these methods cover a large portion of parameter space described by Lindblad equations in quantum optics. However in the few multi-level system limit, with strong correlations and systems outside the few excitation limit these methods are not well suited. For these applications we believe the permutation symmetry based method of PsiQuaSP may be advantageous compared to existing methods, not only because it is a non-approximate scheme. However the exact method presented in this report can also be used to test the range of validity of other approximate methods.\\

\subsection{Some theoretical details} 
Exploiting the permutational symmetry of Lindblad equations results in a polynomial complexity in the number of multi-level systems instead of an exponential complexity. This is equivalent to projecting the master equation onto a subspace of special symmetrized Liouville space states. This approach is only valid if the master equation obeys the permutation symmetry. The permutation symmetric states have been introduced and discussed for two-level systems by various authors \cite{Sarkar:JPhysA:87,Sarkar:EuroPhysLett:87,Carmichael::02,Hartmann:QIC:16,Xu:PhysRevA:13,Richter:PhysRevB:15,Gegg:NJP:16,Kirton:PhysRevLett:17,Gegg:arxiv:17}, notably Hartmann called them generalized Dicke states \cite{Hartmann:QIC:16}. For multi-level systems the scheme can be derived by explicitly looking at the time evolution of density matrix elements (see Ref. \citenum{Gegg:NJP:16} and \citenum{Richter:PhysRevB:15}).\\
For a collection of $d+1$-level systems these special symmetrized Liouville space states are given by
\begin{equation}
\hat{\mathcal{P}}[\{n_{kl}\}] = \mathcal{S} \bigotimes_{k,l=0}^{d} \sigma_{kl}^{\otimes n_{kl}},
\label{eq.defbasis}
\end{equation}
where $\{n_{kl}\} = \{n_{dd}, n_{d(d-1)}, \dots\}$ is the set of all numbers $n_{kl}$. The product in equation \eqref{eq.defbasis} consists of $N$ individual spin matrices, one for each multi-level system. Thus in this direct product each spin is exactly represented by one of the $(d+1)^2$ individual spin matrices and $n_{kl}$ spins are in the same $\sigma_{kl}$ state. The ordering in this product is not uniquely determined, there are many permutations that can be written as such a product of spin matrices, characterized by the numbers $\{n_{kl}\}$. The symmetrization operator $\mathcal{S}$
\begin{equation}
\mathcal{S} = \sum_P \hat{P}
\end{equation}
creates a sum over all these possible permutations $P$ of the spin matrices $\sigma_{kl}^i$ for a given configuration $\{n_{kl}\}$. Here $\hat{P}$ is the permutation operator. This results in an unambiguous definition of totally symmetrized states. Please note that our definition of the symmetrization operator does not contain a normalization factor in contrast to the symmetrization operator usually used for constructing $N$ particle boson states. Omitting the normalization makes the method numerically more stable, see Ref. \citenum{Gegg:NJP:16}.\\
The number of possible permutations is given by a multinomial coefficient
\begin{equation}
{N \choose \{n_{kl}\}} = \frac{N!}{n_{dd}! n_{d(d-1)}! \dots n_{00}!}.
\label{eq.defmultinom}
\end{equation} 
This can be justified as follows: A set of $N$ multi-level systems is divided into $(d+1)^2$ subsets, one for each individual spin matrix. Then the $n_{kl}$ are the numbers of elements in these sets and the number of possible realizations is given by the multinomial coefficient equation \eqref{eq.defmultinom}. Please note that this is precisely why this method has a polynomial instead of exponential complexity: Each density matrix element corresponding to one of the permutations in equation \eqref{eq.defbasis} is identical to the density matrix elements of all the other permutations. This holds if the master equation has permutation symmetry and the system is prepared in an initial state that obeys permutation symmetry. This requirement is fulfilled if the system is prepared in e.g. the ground or a thermal equilibrium state. 
Summing over all states that correspond to these identical density matrix elements results in equation \eqref{eq.defbasis}.\\ 
The product contains exactly one spin matrix per multi-level system, this implies
\begin{equation}
\underbrace{\sum_{ij} n_{kl}}_{m~\mbox{summands}} = N, \quad \mbox{with} \quad 0 \leq n_{kl}.
\label{eq.sourceofcomplexity}
\end{equation}
This expression determines the number of different basis states and thus the complexity or dimensionality of the problem: The number of possible sets $\{n_{kl}\}$ that satisfy equation \eqref{eq.sourceofcomplexity} is given by\cite{Gegg:NJP:16}
\begin{equation}
{N + m-1 \choose N} \propto \frac{N^{m-1}}{(m-1)!},
\label{eq.defcomplexity}
\end{equation}
hence the method scales polynomially, with the order of the polynom depending on the number $m$ of different numbers $n_{kl}$. Please note that the number of involved spin matrices does not have to be identical to $(d+1)^2$, the total number of individual spin matrices for a $d+1$-level system. It can be lower if additional symmetries apply (see below).\\
The basis states defined in equation \eqref{eq.defbasis} are orthogonal with respect to the Hilbert-Schmidt inner product. See Ref. \citenum{Gegg:NJP:16} for a detailed explanation. equation \eqref{eq.sourceofcomplexity} allows to eliminate one of the $n_{kl}$ coefficients. We usually eliminate $n_{00}$, the number of multi-level systems sitting in the ground state.\\
As an illustration we consider two two-level systems: The permutation symmetric two-level system states are described by three numbers $n_{11}$, $n_{10}$, $n_{01}$ (omitting $n_{00}$), the basis elements are
\begin{equation}
\hat{\mathcal{P}}[n_{11},n_{10},n_{01}] = \mathcal{S} ~\sigma_{11}^{\otimes n_{11}} \sigma_{10}^{\otimes n_{10}} \sigma_{01}^{\otimes n_{01}} \sigma_{00}^{\otimes n_{00}},
\end{equation}
where $n_{00}=N-n_{11}-n_{10}-n_{01}$.
\begin{table}
\centering
\begin{tabular}{lclc}
$\hat{\mathcal{P}}[0,0,0]=$&$\sigma_{00}^1\sigma_{00}^2$& $\hat{\mathcal{P}}[1,0,0] =$ & $\sigma_{11}^1\sigma_{00}^2 + \sigma_{00}^1\sigma_{11}^2$\\
$\hat{\mathcal{P}}[2,0,0]=$&$\sigma_{11}^1\sigma_{11}^2$& $\hat{\mathcal{P}}[0,1,0] =$ & $\sigma_{10}^1\sigma_{00}^2+\sigma_{00}^1\sigma_{10}^2$\\
$\hat{\mathcal{P}}[1,1,0]=$&$\sigma_{11}^1\sigma_{10}^2 + \sigma_{10}^1\sigma_{11}^2$ & $\hat{\mathcal{P}}[0,2,0] =$ & $\sigma_{10}^1\sigma_{10}^2$\\
$\hat{\mathcal{P}}[0,0,1]=$&$\sigma_{01}^1\sigma_{00}^2+\sigma_{00}^1\sigma_{01}^2$&$\hat{\mathcal{P}}[1,0,1]=$&$\sigma_{11}^1\sigma_{01}^2+\sigma_{01}^1\sigma_{11}^2$\\
$\hat{\mathcal{P}}[0,1,1]=$&$\sigma_{10}^1\sigma_{01}^2+\sigma_{01}^1\sigma_{10}^2$&$\hat{\mathcal{P}}[0,0,2]=$&$\sigma_{01}^1\sigma_{01}^2$ 
\end{tabular}
\caption{All permutational symmetric basis states for $2$ two-level systems. Swapping the indices $1 \leftrightarrow 2$ leaves these states invariant. The actions of all terms in a permutationally symmetric master equation only connects these $10$ basis states. This is an instructive example for understanding the basis, the reduction in complexity however is minimal for this case ($10$ states compared to $2^{2\cdot2} = 16$ states for the full approach).}
\label{tab.basistwotwo}
\end{table}
According to equation \eqref{eq.defcomplexity} this results in ${2+3 \choose 2} = 10$ possible basis states. These $10$ states are given in Table \ref{tab.basistwotwo}.\\
The $\hat{\mathcal{P}}[\{n_{kl}\}]$ notation is the formal foundation for PsiQuaSP, but it is not very intuitive and not useful for setting up a simulation. Therefore in Ref. \citenum{Gegg:NJP:16} we have developed a sketch representation for these elements. PsiQuaSP is designed in a way that allows the user to translate the sketches directly into code. The user does not need to derive any equations of motion, which facilitates the usage and greatly speeds up code development time. However a basic understanding of the sketches as well as the principles of the permutation symmetry is crucial for a successful usage of PsiQuaSP.\\
The sketches are intended to visualize the physical processes associated with the different contributions in the master equation: Looking at the time evolution of the respective density matrix elements
\begin{equation}
\mathbf{tr}[\hat{\mathcal{P}}[\{n_{kl}\}] \rho] = \mathcal{P}[\{n_{kl}\}],
\label{eq.defelements}
\end{equation}
we can e.g. derive the two-level system spontaneous emission contribution equation \eqref{eq.tlsspontem}
\begin{align}
\partial_t \mathcal{P}[n_{11},n_{10},n_{01}]\big\vert_{\mathcal{D}_1(\rho)} &= \frac{\gamma}{2} \bigg[2(n_{11}+1) \mathcal{P}[n_{11}+1,n_{10},n_{01}]\nonumber\\
& \quad - (2 n_{11} +n_{10}+n_{01})\mathcal{P}[n_{11},n_{10},n_{01}]\bigg].
\end{align}
This equation describes density decay as well as decay induced dephasing of quantum coherences. \\
The density matrix elements $\mathcal{P}[n_{11},n_{10},n_{01}]$ correspond to a quantum coherence/correlation for $n_{10},n_{01} \neq 0$, for $n_{10},n_{01} = 0$ the element corresponds to the population of the $N$ two-level system setup in a state with $n_{11}$ excitations, a density. Generally for elements $\mathcal{P}[\{n_{kl}\}]$, if the numbers $n_{kl}$ corresponding to flip operators ($i\neq j$) are zero, then the element is a density, otherwise the element corresponds to a quantum coherence.\\
We visualize the decay process as arrows by drawing the four degrees of freedom $n_{00}$, $n_{01}$, $n_{10}$ and $n_{11}$ as bubbles, see Fig. \ref{fig.mlsintro} a). The full sketch for the master equation for the Dicke Hamiltonian equation \eqref{eq.dickeham} and the two-level system spontaneous emission is shown in Fig. \ref{fig.mlsintro} b). Fig. \ref{fig.mlsintro} c) shows the sketch for the $\Lambda$ setup. Fig. \ref{fig.mlsintro} d) corresponds to a three-level laser setup, which has further symmetries that lead to an additional reduction in degrees of freedom and thus also dimensionality/numerical complexity. 
For the three-level laser setup $4$ coherence degrees of freedom ($n_{20}$, $n_{21}$, $n_{02}$ and $n_{12}$) are decoupled from the densities. Since these decoupled coherences only dephase and are not driven we can set them to zero. More specific, for every basis element in equation \eqref{eq.defbasis} with a nonzero index $n_{20}$, $n_{21}$, $n_{02}$ or $n_{12}$ the corresponding density matrix element is always zero. For more information on the rules for constructing the sketches please refer to Ref. \citenum{Gegg:NJP:16} and Section \ref{sec.sketches}.\\
PsiQuaSP uses PETSc or dependent packages to compute all relevant density matrix elements
\begin{equation}
\mathbf{tr}[\hat{\mathcal{P}}[\{n_{kl}\}] \rho],
\label{eq.defhilberschmidtexpansion}
\end{equation}
which are represented by a single column vector in memory. The Liouville operator $\mathcal{L}$ is thus represented as a matrix. PsiQuaSP provides all functionality to setup an arbitrary master equation and observables, etc. that is compatible with the permutation symmetry scheme. The translation into the internal representation used for calculation is completely handled by PsiQuaSP. As a free bonus also master equations without permutation symmetry can be set up, since any system described by a finite dimensional Hilbert space can be represented as a single multi-level system.

\section{Using PsiQuaSP -- Basic structure of the library}
\label{sec.basicst}

\begin{figure}
\centering
\includegraphics[scale=0.3]{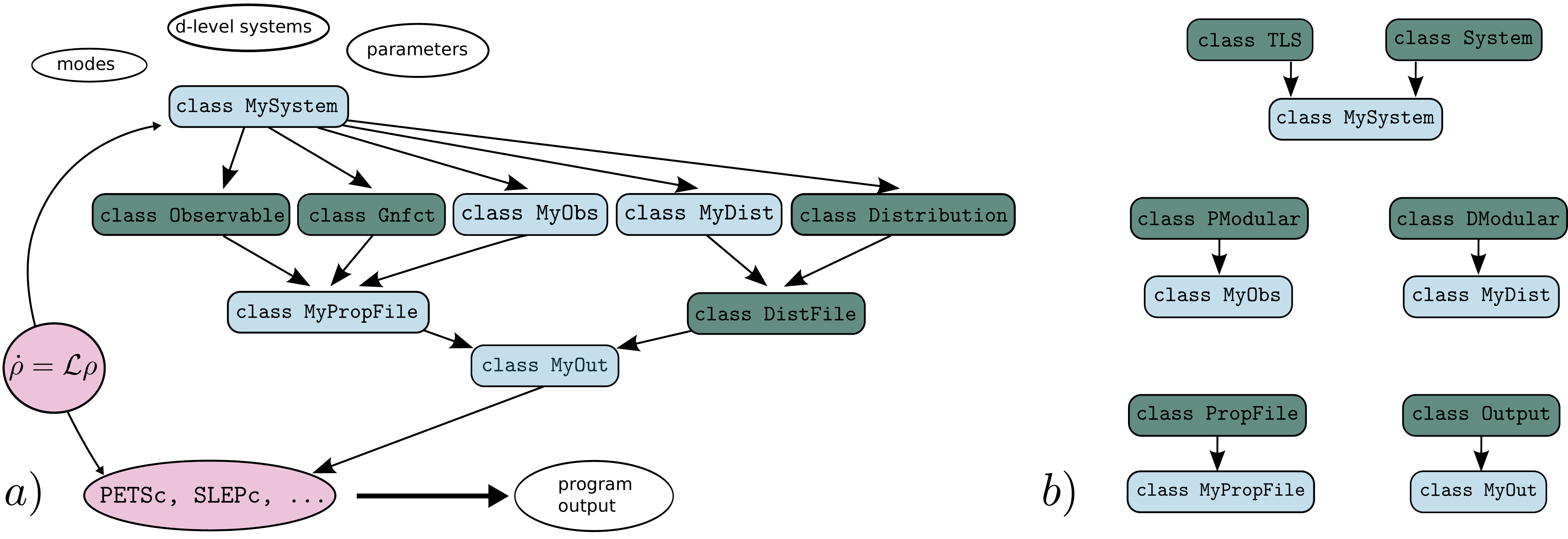}
\caption{a) Schematic representation of the general structure of a PsiQuaSP application code: \texttt{MySystem} contains all the relevant information about the system and is used to construct the master equation and the output. The output is organized in three layers, the first layer consists of objects that can compute the desired properties of the system, like \texttt{Observable}, \texttt{Distribution}, the correlation functions \texttt{Gnfct} and the custom types \texttt{MyObs} and \texttt{MyDist}. The second layer groups these objects into output files, each managed by another object. The third layer consists of the \texttt{MyOut} class, which groups all output files and provides a clean interface to PETSc. Classes that need to be derived from base classes have blue boxes, green boxes indicate ready to use classes. b) Base class diagram for the derived classes in a). Only for \texttt{MySystem} there are two possibilities: \texttt{TLS} for two-level system setups and \texttt{System} for all other purposes.}
\label{fig.pqspstruct}
\end{figure}

PsiQuaSP is designed in a way that provides maximal flexibility for setting up simulations. Therefore PsiQuaSP provides setup routines, i.e. for constructing the density matrix and the Liouvillians $\mathcal{L}$ and it allows to define observables, distributions, correlation functions etc. and encapsulates them in a user friendly way. The numerical solution solely relies on PETSc, derived packages such as SLEPc and/or external packages that can be used with PETSc such as MUMPS, SuperLU, Metis/ParMetis, PTScotch and others\cite{petsc-web-page,petsc-user-ref,petsc-efficient,Hernandez:LNCS:03,Hernandez:TMS:05,slepc-users-manual,mumps,Li:TMS:05,parmetis,Chevalier:PC:08,ptscotch}, like in other PETSc based libraries\cite{quac}. All these packages are incredibly valuable and powerful tools and PETSc provides a unified interface to all these packages. This is one of the reasons that makes PETSc an excellent basis for PsiQuaSP. Getting to know all these packages requires a lot of time and effort, but the average user can use PsiQuaSP without knowing these additional packages. However we wish to encourage the readers of this report to look into these packages and related numerical literature and find out what they can do in order to boost the performance of the numerical code. The right choice of method can reduce computing time by orders of magnitude, see Fig. \ref{fig.perf} (b).\\
The heart of PsiQuaSP is the \texttt{System} class. The user first specifies whether two-, three- or d-level systems are used and how many bosonic modes are required. For two-level systems the special \texttt{TLS} class provides further encapsulation and therefore simplification for standard two-level system Hamiltonians and dissipators. Based on the information on multi-level sytems and bosonic modes the \texttt{System}/\texttt{TLS} class provides initialization functions for the density matrix and Liouvillians -- thus everything required for setting up the master equation. The output of the program is managed by the \texttt{Output} class, which can manage a set of user defined output files, containing observables, correlation functions, distributions, etc, see Fig. \ref{fig.pqspstruct} a). Please note that even though PsiQuaSP is intended and designed for solving permutationally symmetric master equations, the library is not limited to this application. It may also be used for efficient treatments of nonidentical multi-level systems as well as Hamiltonian diagonalizations.\\
Installation instructions for PsiQuaSP and PETSc are given in the \texttt{README.md} and \texttt{INSTALL.md} files in the PsiQuaSP folder. PsiQuaSP uses Doxygen commenting. Doxygen translates the comments in the source code into a structured website representation, which is extremely useful for getting to know the library. Read \texttt{doc/README.md} for further information.\\
We will now illustrate the setup of PsiQuaSP simulations along the discussion of some examples. All source codes of the examples and many more can be found in the \texttt{example/} directory in the PsiQuaSP directory.
\subsection*{Example 1: Open Tavis-Cummings relaxation} 
The first example is the master equation defined by equations \eqref{eq.tlsmaster}, \eqref{eq.dickeham} and \eqref{eq.tlsspontem} (Figs. \ref{fig.mlsintro} a) and b)). This is a basic Tavis-Cummings/Dicke model including individual spontaneous decay of the two-level systems and a cavity loss term. The example code computes the temporal dynamics of this master equation using direct Runge-Kutta time integration. The source code can be found in \texttt{example/ex1a}. \texttt{example/ex1b} solves the same equation with an adaptive step width Runge-Kutta and at the same time shows the application of more advanced PETSc routines. Since there is no pump term in this master equation the steady state is the ground state and we need to prepare the system initially in an excited state in order to observe nontrivial dynamics.\\
\emph{System/Master equation setup:} First we declare a derived class for the system under consideration:

\texttt{class OTC: public TLS}

\texttt{\{}

\texttt{  public:}

\texttt{    void	Setup(Vec * dm, Mat * L);}

\texttt{\};}\\
This class just defines a setup function. This is the standard procedure in PsiQuaSP, i.e. in most cases user derived classes just define a setup function. We use the base class TLS, which provides enhanced tools for master equations only involving two-level systems. Here the setup function will create a vector \texttt{Vec * dm} and a matrix \texttt{Mat * L}, which are the density matrix and the Liouvillian of the system. PsiQuaSP uses a vectorized version of the master equation. The two types \texttt{Vec} and \texttt{Mat} are defined by PETSc. Both can be either serial or parallel, \texttt{Mat} is sparse by default (but dense types are available if needed), leading to efficient memory usage and reduction in computation time.\\
In the \texttt{OTC::Setup(...)} function we call the functions

\texttt{TLSAdd(ntls,ntls,ntls,tlsenergy);}

\texttt{ModeAdd(m0+1,dm0,modeenergy);}

\texttt{PQSPSetup(dm,1,L);}\\
to tell PsiQuaSP that we are considering \texttt{nlts} two-level systems and one bosonic mode with maximum Fock state \texttt{m0}. \texttt{TLSAdd(...)} adds the two-level system quantum numbers $n_{11}$, $n_{10}$ and $n_{01}$, c.f. Fig. \ref{fig.mlsintro} a) and b). The three arguments \texttt{ntls,ntls,ntls} specify the maximum number for the three indices $n_{11}$, $n_{10}$, $n_{01}$. This allows a truncation of the three individual quantum numbers. \texttt{tlsenergy} and \texttt{modeenergy} are the transition energies for exciting a two-level system and the photon energy, which are needed for preparing the system in a thermal equilibrium state. These energy parameters are independent of the parameters used for the equation of motion since a rotating frame representations might be used. These energy parameters are usually written into the file headers and are needed for thermal state preparation and have no other purpose! After this the user needs to call \texttt{PQSPSetup()}, the setup function for all internal structures which creates the density matrix vector \texttt{dm} and the Liouvillian matrix \texttt{L}. Now the master equation is specified. This is done by calling

\texttt{AddTLSH0(*L,NULL,NULL,1,domega\_tls*PETSC\_i);}

\texttt{AddTavisCummingsHamiltonianRWA(*L,NULL,NULL,1,0,gcouple*PETSC\_i);}

\texttt{AddTLSSpontaneousEmission(*L,NULL,NULL,1,gamma/2.0);}

\texttt{AddLindbladMode(*L,NULL,NULL,1,0,kappa/2.0);}\\
Here each line adds the contributions of a different term of the master equation to the Liouvillian matrix \texttt{L}. The sketch for \texttt{AddTLSSpontaneousEmission(..)} is shown in Fig. \ref{fig.mlsintro} a) and \texttt{AddTavisCummingsHamiltonianRWA(..)} is represented by the green arrows in Fig. \ref{fig.mlsintro} b). \\
Mode related Liouvillians like \texttt{AddLindbladMode(...)} are not represented with sketches and the \texttt{AddTLSH0(..)} is given by the combination of sketches Fig. \ref{fig.modularexample} a) and b). In this example we use a rotating frame representation and $\mbox{\texttt{domega\_tls}}$ is the detuning of the two-level systems from the cavity mode, on resonance $\mbox{\texttt{domega\_tls}}=0.0$ holds. The next step is to specify initial conditions, here we prepare the system in an excited state:

\texttt{PetscInt	qnumbers [5] = \{n11,n10,n01,mket,mbra\};}

\texttt{DMWritePureState(*dm,qnumbers);}\\
The \texttt{qnumbers} array contains the quantum numbers of the desired state. This setup function can prepare the density matrix in any of the permutation symmetric basis states equation \eqref{eq.defbasis}. However only for $\mbox{\texttt{n10}}= \mbox{\texttt{n01}} = 0$ and $\mbox{\texttt{mket}}= \mbox{\texttt{mbra}}$ the state corresponds to a physically meaningful population. This function addresses the different quantum numbers in the order they have been set: As stated above the \texttt{TLSAdd(...)} function call adds the two-level system quantum numbers in the order $n_{11}$, $n_{10}$, $n_{01}$ and the function \texttt{ModeAdd(...)} always adds first the ket and then the bra quantum number of $|\mbox{\texttt{mket}}\rangle \langle \mbox{\texttt{mbra}}|$. If we had added another mode via two successive calls to \texttt{ModeAdd(...)} we would address an individual state with an array like 

\texttt{PetscInt	qnumbers [7] = \{n11,n10,n01,m0ket,m0bra,m1ket,m1bra\};}\\
PsiQuaSP internally labels the modes with numbers starting from $0$ in the order of creation.\\
To create an object of the system specification class \texttt{OTC} that we just defined, we call e.g. in the main routine

\texttt{OTC	otc;}

\texttt{otc.Setup(\&dm,\&L);}\\
The \texttt{otc} object has two purposes: It creates all ingredients to the master equation and after successful setup it contains all necessary information about the system. Afterwards the object is used to build observables and to specify the output data.\\
\emph{Defining the output:} The expectation value of a collective operator like $J_{11}$ (equation \eqref{eq.defcollspinop}), which represents the mean occupation of the excited states of all two-level systems
\begin{equation}
\langle J_{11} \rangle = \mathbf{tr}[J_{11}\rho] = \sum_{n=0}^N \sum_m n \mathcal{P}[n,0,0;m,m]
\end{equation}
can be defined with the command

\texttt{Observable	*pdens11	= new Observable();}

\texttt{MLSDim	n11 (1,1);}

\texttt{pdens11->SetupMlsOccupation(otc,n11);}\\
Here \texttt{n11} is an identifier referring to the $n_{11}$ degree of freedom and the function \texttt{SetupMlsOccupation()} can define all $\langle J_{kk} \rangle$ observables for arbitrary multi-level systems
\begin{equation}
\langle J_{kk} \rangle =  \sum_{n_{kk}=0}^N \sum_{\dots} n_{kk} \mathcal{P}[\dots n_{kk}\dots],
\end{equation}
where the second sum runs over all indices describing a density, e.g. a partial trace. The \texttt{MLSDim} and \texttt{ModeDim} classes provide a way to access different degrees of freedom within the application code. Output files that print observables, distributions etc. at every $n$th time step are also managed by classes,  $n$ is equal to $30$ by default and can be changed with the \texttt{-tev\_steps\_monitor newvalue} command line option. For files printing observables like $J_{kk}$ the user creates a derived class like

\texttt{class ObservablesFile: public PropFile}

\texttt{\{}

\texttt{  public:}

\texttt{    void	SetupMyObsFile(OTC * otc, std::string name);}

\texttt{\};}\\
As in the \texttt{OTC} class only the definition of a setup function is required. \texttt{name} is the name for the output file. This class is derived from the \texttt{PropFile} class. Classes derived from this class allow the user to print an arbitrary number of user specified properties that are related to operator expectation values. This includes standard (already implemented) observables of the \texttt{Observable} class, correlation functions $g^{(n)}(\tau)$ (\texttt{Gnfct} class) and user defined custom observables (\texttt{PModular} class). Within this setup function we create the \texttt{Observable} object as above and add it to the output file with the command

\texttt{AddElem(pdens11,"<J\_11>");}\\
The second argument is the name of this quantity in the header of the file. This observables file including an arbitrary number of other user specified output files is bundled into the \texttt{MyOut} class

\texttt{class MyOut: public Output}

\texttt{\{}

\texttt{  public:}

\texttt{    void	SetupMyOut(OTC * system);}

\texttt{\};}\\
The setup function includes the following function calls

\texttt{ObservablesFile	*obsfile	= new ObservablesFile;}

\texttt{obsfile->SetupMyObsFile(system,"observables.dat");}

\texttt{AddOFile(obsfile);}\\
We can specify an arbitrary number of different output files for customized purposes by either providing multiple setup functions in \texttt{ObservableFile} class or deriving a new class with a single setup function for each file e.g. \texttt{ObservableFile1}, \texttt{ObservableFile2} $\dots$.  The \texttt{DistFile} class is used for number state distributions of the modes and the multi-level systems, as well as more complicated (also custom made) distributions like the \texttt{DickeDistribution}.\\ 
As for the \texttt{OTC} class e.g. in the main file we need to call

\texttt{MyOut	*out = new MyOut;}

\texttt{out->SetupMyOut(\&otc);}\\
These function calls create the whole output structure of the program bundled into one object. Generally PsiQuaSP provides functionality for setting up vectors and matrices and to create the output object (\texttt{out}). These three objects (\texttt{Vec}, \texttt{Mat} and \texttt{MyOut}) then provide the input fed into the PETSc (SLEPc, ...) solution routines. Please note that PETSc vectors are not always density matrices and matrices are not always Liouvillians for the master equation. For example the trace operation and therefore any computation of an observable can be defined as a PETSc vector (vector of dual space). This linear functional is subsequently applied to the density matrix via a scalar product using the PETSc routine \texttt{VecDot(...)}: Defining a custom observable $\langle \hat{O}\rangle$ usually is done by setting the matrix for the Liouvillian corresponding to the action of $\hat{O} \rho$, multiplying it with the trace vector and storing the resulting vector/linear functional. The computation of the observable is then calculated using the scalar product of this vector with the density matrix vector (\texttt{VecDot(...)}). This is shown in \texttt{example/ex2a}.\\
\emph{Solving with PETSc:} The numerical solution is handled by PETSc, in this example we use simple time integration using a normal fourth order Runge-Kutta in \texttt{example/ex1a} and an adaptive time step Runge-Kutta in \texttt{example/ex1b}. The basic setup of a time integration using PETSc is as follows:

\texttt{TS		ts;}

\texttt{TSCreate(PETSC\_COMM\_WORLD,\&ts);}

\texttt{TSSetType(ts,TSRK);}\\
This creates the PETSc time stepper context \texttt{TS} and sets it to Runge-Kutta. \texttt{PETSC\_COMM\_WORLD} is the PETSc MPI communicator. PsiQuaSP is fully parallelized by default by using the PETSc routines, but it can of course always be run on a single processor. With the commands

\texttt{TSSetRHSFunction(ts,NULL,TSComputeRHSFunctionLinear,NULL);}

\texttt{TSSetRHSJacobian(ts,L,L,TSComputeRHSJacobianConstant,NULL);}\\
we tell PETSc that the right hand side of the differential equation ($\dot{\rho} = \mathcal{L} \rho$) is given by a constant matrix and that this matrix is the \texttt{L} matrix. The output of the PETSc time steppers is handled by a monitor function, which is a function with a defined interface that PETSc calls at every integration step:

\texttt{TSMonitorSet(ts,MyOut::TEVMonitor,out,NULL);}\\
The function \texttt{MyOut::TEVMonitor} is the monitor function for time integration in PsiQuaSP. It prints a single line at every $n$th time step into each specified output file by computing all user specified observables and distributions in each individual file (\texttt{-tev\_steps\_monitor newvalue} to change $n$). The command

\begin{figure}
\centering
\includegraphics[scale=0.8]{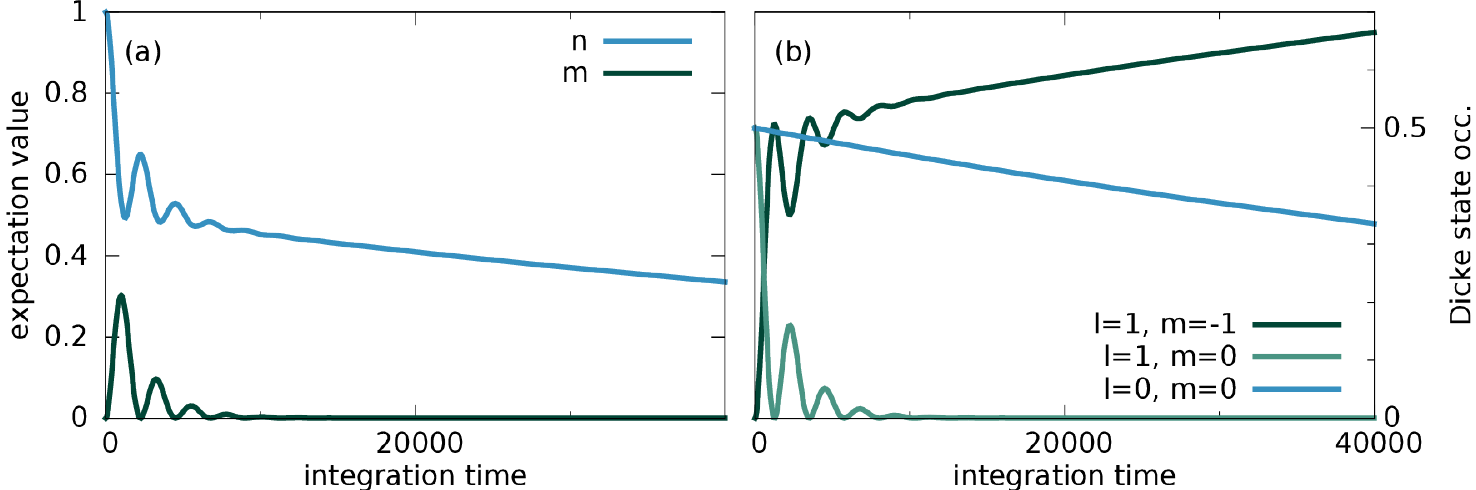}
\caption{Using the code of \texttt{example/ex1b}: a) mean excitation in the two-level systems $n=\langle J_{11}\rangle$ and mean photon number $m=\langle b^\dagger b \rangle$ for $2$ two-level systems prepared in the state $\mathcal{P}[1,0,0;0,0]$. This corresponds to the entanglement distillation setup\cite{Tanas:JOptB:04}. The bright superradiant states couple to the cavity mode and cause Rabi oscillations, while the dark subradiant state does not couple to the cavity and just decays via individual spontaneous emission\cite{Mandel::95,Gegg:arxiv:17}, c.f. equation \eqref{eq.tlsspontem}. b) Dicke state occupations $\langle |l,m\rangle \langle l,m |\rangle$: temporal dynamics of the states of the superradiant subspace (green) vs the single dark state in the subradiant subspace (blue).}
\label{fig.tlsexample}
\end{figure}

\texttt{TSSolve(ts,dm);}\\
solves the time dynamics, \texttt{dm} contains always the current time step density matrix. In Fig. \ref{fig.tlsexample} (a) the mean excitation in the two-level systems and the mode during this relaxation is shown: Initially the dynamics are fast due to Rabi oscillations between bright Dicke states and the mode. Afterwards the dynamics is governed by the slow, monotonous spontaneous emission, since only the dark Dicke states remain excited. In Fig. \ref{fig.tlsexample} (b) the population in these Dicke states is shown.
\subsection*{Example 2: Three-level systems:} 
In the two-level system example above we used the base class \texttt{TLS}. For three- and general multi-level systems specialized classes are not provided, instead there is the multi purpose class \texttt{System} (the base class of \texttt{TLS}). In Figs. \ref{fig.mlsintro} c) and d) two different three-level system sketches are shown: Fig. \ref{fig.mlsintro} c) connects all degrees of freedom while in Fig. \ref{fig.mlsintro} d) four degrees of freedom can be eliminated, resulting in a $\sim N^4$ scaling instead of an $\sim N^8$ scaling for Fig. \ref{fig.mlsintro} c). The decoupling of some basis states and the resulting reduction in degrees of freedom is the main reason why PsiQuaSP does not provide specialized classes for multi-level systems.\\
For the two-level system example we called \texttt{TLSAdd(a,b,c,energy)} which uses internally

\texttt{MLSAddDens(n11,a+1,energy);}

\texttt{MLSAddPol(n10,b+1);}

\texttt{MLSAddPol(n01,c+1);}\\
where \texttt{MLSAddDens(...)} adds a density degree of freedom, corresponding to a quantum number $n_{xx}$, and \texttt{MLSAddPol(...)} adds a polarization degree of freedom, corresponding to a quantum number $n_{xy}, ~x\neq y$. Thus the degrees of freedom for three level systems (Fig. \ref{fig.mlsintro} c)) are set with the function calls (without truncating basis states)

\texttt{MLSAddDens(n22,n+1,energy2);}

\texttt{MLSAddPol(n21,n+1);}

\texttt{MLSAddPol(n20,n+1);}

\texttt{MLSAddPol(n12,n+1);}

\texttt{MLSAddDens(n11,n+1,energy1);}

\texttt{MLSAddPol(n10,n+1);}

\texttt{MLSAddPol(n02,n+1);}

\texttt{MLSAddPol(n01,n+1);}\\
or for Fig. \ref{fig.mlsintro} d) with

\texttt{MLSAddDens(n22,n+1,energy2);}

\texttt{MLSAddDens(n11,n+1,energy1);}

\texttt{MLSAddPol(n10,n+1);}

\texttt{MLSAddPol(n01,n+1);}\\
Here \texttt{n} represents the number of considered three-level systems. This number can also be lower than the number of treated three-level systems, which corresponds to a truncation of the number of three-level system basis states. A truncation should always be tested if it is applicable in the given situation (parameter dependent), but it can reduce the numerical cost considerably (Example: strong dephasing in driven systems can reduce the number of needed offdiagonals (\texttt{nxy}) considerably). The \texttt{nxy} objects are again the \texttt{MLSDim} identifiers and are created with e.g. 

\texttt{MLSDim	n21 (2,1);}\\
As in the two-level system example, after setting all multi-level system degrees of freedom the user adds bosonic modes with the command

\texttt{ModeAdd(m0+1,dm0,modeenergy);}\\
This order of \texttt{ModeAdd(...)} calls after the \texttt{MLSAdd...} calls is mandatory, PsiQuaSP returns an error message if these routines are not called in the right order. Setting e.g. the spontaneous emission dissipator between levels $1 - 0$ for Fig. \ref{fig.mlsintro} c) is done with

\texttt{AddLindbladRelaxMLS(L,NULL,NULL,1,n11,n00,gamma/2.0);}

\texttt{AddLindbladDephMLS(L,NULL,NULL,1,n10,gamma/2.0);}

\texttt{AddLindbladDephMLS(L,NULL,NULL,1,n01,gamma/2.0);}

\texttt{AddLindbladDephMLS(L,NULL,NULL,1,n21,gamma/2.0);}

\texttt{AddLindbladDephMLS(L,NULL,NULL,1,n12,gamma/2.0);}\\
and for Fig. \ref{fig.mlsintro} d) it is

\texttt{AddLindbladRelaxMLS(L,NULL,NULL,1,n11,n00,gamma/2.0);}

\texttt{AddLindbladDephMLS(L,NULL,NULL,1,n10,gamma/2.0);}

\texttt{AddLindbladDephMLS(L,NULL,NULL,1,n01,gamma/2.0);}\\
The parameter \texttt{gamma/2.0} is the same parameter as in equation \eqref{eq.tlsspontem} and each function call corresponds to exactly one arrow in the sketches. Incoherent pumping is added by calling

\texttt{AddLindbladRelaxMLS(L,NULL,NULL,1,n00,n22,pump/2.0);}\\
and the respective calls to \texttt{AddLindbladDephMLS()}. The interaction of the three-level systems with the mode for Fig. \ref{fig.mlsintro} c) is included by calling

\texttt{AddMLSModeInt(AA,NULL,NULL,1,n20,n21,mbra,-gcouple*PETSC\_i);}

\texttt{AddMLSModeInt(AA,NULL,NULL,1,n10,n11,mbra,-gcouple*PETSC\_i);}

\texttt{AddMLSModeInt(AA,NULL,NULL,1,n00,n01,mbra,-gcouple*PETSC\_i);}

\texttt{AddMLSModeInt(AA,NULL,NULL,1,n02,n12,mket,gcouple*PETSC\_i);}

\texttt{AddMLSModeInt(AA,NULL,NULL,1,n01,n11,mket,gcouple*PETSC\_i);}

\texttt{AddMLSModeInt(AA,NULL,NULL,1,n00,n10,mket,gcouple*PETSC\_i);}\\
and for Fig. \ref{fig.mlsintro} c) omitting the arrows of the disconnected part of the sketch

\texttt{AddMLSModeInt(AA,NULL,NULL,1,n10,n11,mbra,-gcouple*PETSC\_i);}

\texttt{AddMLSModeInt(AA,NULL,NULL,1,n00,n01,mbra,-gcouple*PETSC\_i);}

\texttt{AddMLSModeInt(AA,NULL,NULL,1,n01,n11,mket,gcouple*PETSC\_i);}

\texttt{AddMLSModeInt(AA,NULL,NULL,1,n00,n10,mket,gcouple*PETSC\_i);}\\
\texttt{mket} and \texttt{mbra} are the identifiers for the mode degrees of freedom and are created by calling

\texttt{ModeDim		mket (0,photonnumber);}

\texttt{ModeDim		mbra (1,photonnumber);}\\
\texttt{photonnumber} is the index of the mode. Modes are numbered internally starting from zero in the order they are created with an \texttt{AddMode()} call. Hamiltonian contributions that change the right index of the \texttt{MLSDim} and/or act on the bra side of the mode expansion come with a minus sign. This stems from the commutator in the von-Neumann part of the quantum master equation, see section \ref{sec.sketches} for more details.
The generation of the output as well as the solution stage is completely analogous to the two-level system example. Further examples, illustrating other master equations, custom observables, custom distributions, custom Hamiltonians and Liouvillians as well as other solution techniques and advanced, graph theory based reduction of degrees of freedom are provided in the examples in the \texttt{example/} folder as well as in section \ref{sec.sketches}. An overview of the current examples and the concepts explained in them is given in Table \ref{tab.examples}.
\begin{table}
	\centering
	\begin{tabular}{ll}
		\texttt{example/} & System, concepts, techniques\\
		\hline
		\texttt{ex1a} & Open Tavis-Cummings model, simple observables, distributions, time-integration\\
		\texttt{ex1b} & \texttt{ex1a} with thermal bath, PETSc concepts, adaptive time integration, Dicke distribution\\
		\texttt{ex2a} & Two-level laser, incoherent pump, custom observables \\
		\texttt{ex2b} & Direct steady state/null space computation using SLEPc Krylov-Schur algorithm\\
		\texttt{ex2c} & Two-level laser with Non-RWA terms\\
		\texttt{ex3a} & Lambda system setup, multi-level system usage \\
		\texttt{ex3b} & Three-level laser\\
		\texttt{ex4a} & Phononlaser/Lasercooling setup, custom Liouvillians\\
		\texttt{ex5} & \begin{tabular}{@{}l@{}}Same as \texttt{ex3a}, using ParMETIS graph partitioning to exploit $U(1)$ symmetry,\\ leading to a reduction from $N^8$ to $\sim N^7$\end{tabular}\\
	\end{tabular}
	\caption{Overview over the example codes and the concepts explained/introduced in these examples. \texttt{ex2b} requires an additional SLEPc installation and for \texttt{ex5} it is necessary to build PETSc with the \texttt{--download-parmetis} flag. }
	\label{tab.examples}
\end{table}

\section{Template functions versus custom Liouvillians}
\label{sec.overview}

PsiQuaSP has roughly two types of possible usages. The first usage was presented in the previous section: using ready-made functions for setting arrows of common Hamiltonians and Lindblad dissipators. Generally a single function call to one of these functions represents a single arrow in one of the sketches. First the user draws the sketch representation of the master equation and then directly translates the sketch into code. In the case of two-level systems it is even simpler -- a single function call is sufficient to set a Hamiltonian or dissipator contribution.
\begin{table}
\centering
\begin{tabular}{lll}
Liouvillian & \texttt{System} function & Examples\\
\hline
$H= \hbar \omega_0 b^\dagger b $ & \texttt{AddModeH0()} & \texttt{ex3a}\\
$H= \hbar \omega_{xx} J_{xx}$ & \texttt{AddMLSH0()} & \texttt{ex1a}, \texttt{ex1b}\\
$H= \hbar g (J_{xy}+J_{yx})(b^\dagger + b)$ & \texttt{AddMLSModeInt()} & \texttt{ex2c}\\
$H= \hbar g (J_{xy}b^\dagger + J_{yx} b)$ & \texttt{AddMLSModeInt()}& \texttt{ex1a}, \texttt{ex1b}\\
$H= \hbar E (J_{xy}e^{i \omega t}+J_{yx}e^{-i \omega t})$ & \texttt{AddMLSCohDrive()} & \texttt{ex3a}\\
$H= \hbar E (b e^{i \omega t}+b^\dagger e^{-i \omega t})$ & \texttt{AddModeCohDrive()} & \textbf{none}\vspace{0.15cm} \\ 
\begin{tabular}{@{}l@{}}
$\mathcal{D} = \frac{\gamma}{2} \sum_i (2 \sigma_{xy}^i \rho \sigma_{yx}^i - \sigma_{yy}^i \rho - \rho \sigma_{yy}^i)$ \\
$\mathcal{D} = \delta \sum_i (\sigma_{xy}^{z,i} \rho \sigma_{yx}^{z,i} - \rho) $
\end{tabular}
 & 
\begin{tabular}{@{}l@{}}
\texttt{AddLindbladRelaxMLS()} \\
\texttt{AddLindbladDephMLS()} 
\end{tabular}
& \texttt{ex1a}, \texttt{ex1b} \vspace{0.15cm}\\
$\mathcal{D} = \frac{\kappa}{2} (b \rho b^\dagger - b^\dagger b \rho - \rho b^\dagger b)$ & \texttt{AddLindbladMode()} & \texttt{ex1a}\\
\begin{tabular}{@{}c@{}}
$\mathcal{D} = \frac{\kappa}{2}\big( (\bar{m}+1)(b \rho b^\dagger - b^\dagger b \rho - \rho b^\dagger b)$ \\
$ + \bar{m}(b^\dagger \rho b -  b b^\dagger \rho - \rho b b^\dagger) \big) $ 
\end{tabular} & \texttt{AddLindbladModeThermal()} & \texttt{ex1b}
\end{tabular}
\caption{Overview over the general ready-made Liouvillian setup functions of the \texttt{System} class. Please look into the \texttt{TLS} class documentation to see the derived, specialized two-level system functions. The Hamiltonian contributions always refer to the $i/\hbar[\rho, H]$ terms. Using $\sigma_{xy}^{z,i} = 1/2 (\sigma_{xx} - \sigma_{yy})$. }
\label{tab.redymade}
\end{table}
The implemented contributions are shown in Table \ref{tab.redymade}.\\
In the second usage form the user defines elementary Liouville space operators and constructs arbitrary master equations, observables, distributions, etc. from these elementary operators: The permutation symmetric methodology is in principle applicable to \emph{any} permutation symmetric quantum master equation and using the general framework of PsiQuaSP one can solve in principle \emph{any} quantum master equation in a number state representation (there is currently no support for coherent state basis etc.). Since we cannot provide template setup functions for every conceivable Liouvillian matrix there needs to be another, more flexible approach for this: In the second type of usage the user defines elementary Liouville operators, which act like
\begin{equation}
J_{xy} \rho = J_{xy}^L \rho, \qquad   \rho J_{xy} = J_{xy}^R\rho.
\label{eq.defelemliou}
\end{equation}
Here we used the $L,R$ algebra used in e.g. two-dimensional spectroscopy \cite{Harbola:PhysRep:08}: For any Hilbert space operator we define a Liouville space operator by distinguishing whether it acts on the left or right side of the density matrix, i.e. $A\rho = A^L \rho$ and $\rho A = A^R \rho$. As in the first type of usage the setup of these elementary Liouville operators is done by first drawing a sketch for each needed operator and then adding all needed arrows using single function calls. Based on these elementary operators the user defines arbitrary interaction Hamiltonians and dissipators as well as custom observables, distributions, basis transformations etc. For instance using equation \eqref{eq.defelemliou} the definition of a collective spontaneous emission Liouvillian from level $x$ to level $y$ is
\begin{equation}
\mathcal{D}(\rho) = \frac{\Gamma}{2} (J_{yx}\rho J_{xy} - J_{xy} J_{yx} \rho -  \rho J_{xy} J_{yx}) ~ = ~ \frac{\Gamma}{2} (J_{yx}^L \cdot J_{xy}^R - J_{xy}^L \cdot J_{yx}^L - J_{yx}^R \cdot J_{xy}^R)\rho,
\end{equation}
here the combination of the $R/L$ operators $\cdot$ is performed by the standard matrix-matrix product, provided by the PETSc function \texttt{MatMatMult()}. Hence the user first defines elementary matrices and then uses the PETSc matrix multiplication and addition tools to construct every conceivable Liouville operator. The details for this type of application are explained in the next section.

\section{Building arbitrary Liouvillians}
\label{sec.sketches}

\begin{figure}
\centering
\includegraphics[scale=0.6]{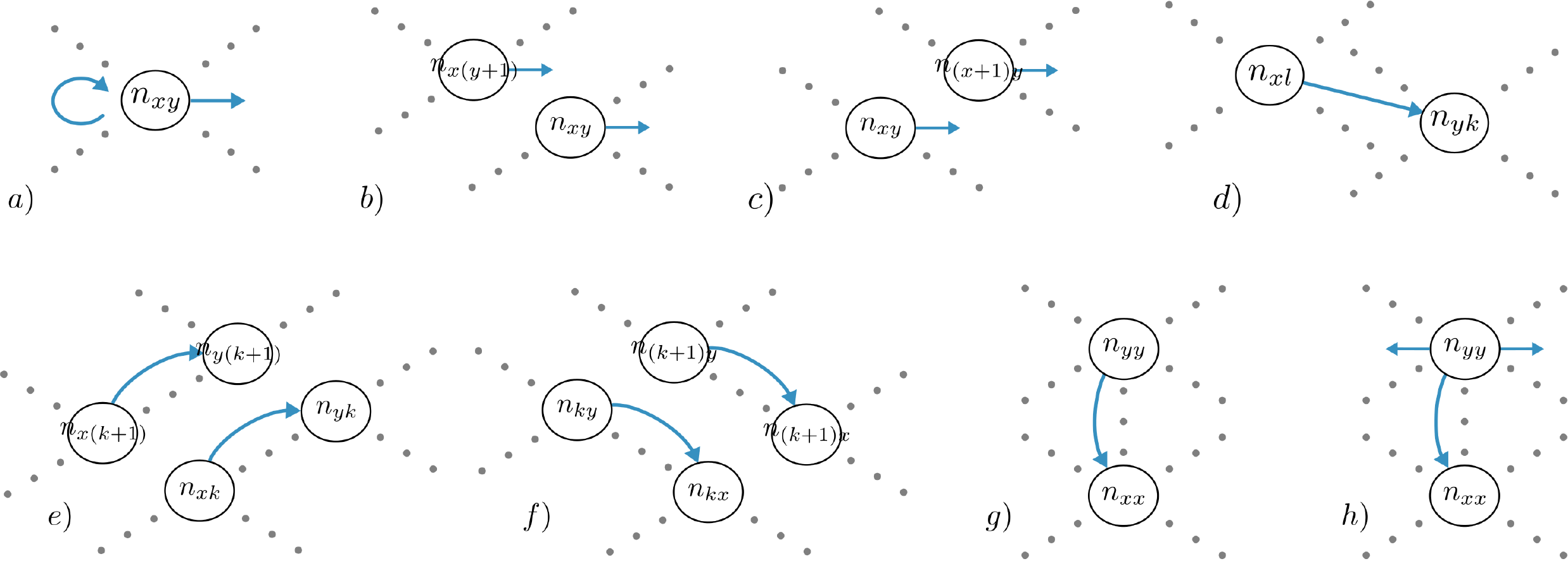}
\caption{Modular sketches for multi-level systems: a) The nonconnecting arrow can represent the phase oscillations arising from the self energy Hamiltonians (curved arrow) and it can describe dephasing (straight arrow). b) and c) the sketches corresponding to dephasing $\dot{\rho} \sim \rho J_{xx}$ and $\dot{\rho} \sim J_{yy} \rho $. d) The connecting arrow can represent flip operators and density relaxation. e) and f) the arrows corresponding to the flip operators $\dot{\rho} \sim \rho J_{xy}$ and $\dot{\rho} \sim J_{xy} \rho $, c.f. equations \eqref{eq.defrightflipdef} and \eqref{eq.defleftflipdef}. g) The density relaxing arrow caused by an individual spontaneous emission like dissipator $\dot{\rho} \sim \sum_i \sigma_{xy}^i\rho\sigma_{yx}^i$. h) The density relaxation arrow introduced in Fig. \ref{fig.mlsintro} a) called by the function \texttt{AddLindbladRelaxMLS()} consists of three arrows in the elementary picture, two nonconnecting and one connecting arrow.}
\label{fig.unamsketchintro}
\end{figure}

In this section we discuss a formalism for the setup of all possible Liouvillians which are consistent with the permutation symmetric method. We introduce separate setup functions for multi-level system and mode degrees of freedom, e.g. for $J_{xy}^L$ or $b^R$. These elementary matrices can be used to construct more complicated operators such as $J_{xy}^L+J_{yx}^R$ and $J_{xy}^L b^L$ by using the PETSc tools for matrix multiplication and addition. Defining such setup functions for the mode degrees of freedom is straightforward and is based on textbook physics\cite{Mukamel::95}. For the symmetric basis states of PsiQuaSP the treatment is a bit more difficult. The following discussion is technical since we want to explain the formal theory underlying PsiQuaSP. The usage however is very simple, it again results in drawing simple sketches and directly implementing arrows by single function calls.\\
\emph{Technical details:} As defined in equation \eqref{eq.defhilberschmidtexpansion} PsiQuaSP uses an expansion of the density matrix in Liouville space. Expansion coefficients are calculated via the Hilbert-Schmidt inner product
\begin{equation}
\mathcal{P}[\{n_{kl}\}] = \mathbf{tr}[\hat{\mathcal{P}}[\{n_{kl}\}]\rho]
\end{equation}
The actions of any operators $A$, $B$ on the density matrix $A\rho B$ is handled by PsiQuaSP like applying these operators to $\hat{\mathcal{P}}[\{n_{kl}\}]$:
\begin{equation}
\mathbf{tr}[\hat{\mathcal{P}}[\{n_{kl}\}] A \rho B]  = \mathbf{tr}[B \hat{\mathcal{P}}[\{n_{kl}\}] A \rho].
\label{eq.rowdef}
\end{equation}
Therefore we introduce a general recipe to construct arbitrary operators $B \hat{\mathcal{P}}[\{n_{kl}\}] A$ expressed in the permutation symmetric basis, for all $A \rho B$ that live in the permutation symmetric subspace. Two steps are necessary: First we need to identify the elementary processes/Liouville operators and second we need to determine how to construct physically relevant operators, like e.g. a collective raising operator for a four-level system acting from the left. The permutation symmetry requires to include only processes acting \emph{indistinguishably} on the left and/or right side of the density matrix. These elementary operators should be representable by arrows.\\
\emph{Defining elementary processes/arrows:} Looking at the sketches in Fig. \ref{fig.mlsintro} we already see two general types of arrows: connecting and nonconnecting arrows. A connecting arrow represents a coupling between two different symmetric basis states equation \eqref{eq.defbasis}, corresponding to an in- or out-scattering process, and a nonconnecting arrow just acts on the state itself, leaving it unchanged. This is quite analogous to the actions of the interacting and non-interacting parts of a Hamiltonian acting on a Hilbert space state, or rather the symmetrized basis states equation \eqref{eq.defbasis} are eigenstates of the operators corresponding to the nonconnecting arrows. It turns out that these are the only possible two types. The general mathematical expressions are given by
\begin{equation}
\sum_i \sigma_{xx}^i \hat{\mathcal{P}}[\dots] \sigma_{yy}^i = n_{xy} \hat{\mathcal{P}}[\dots]
\label{eq.defsinglearrownonconnecting}
\end{equation}
for a single nonconnecting arrow and
\begin{equation}
\sum_i \sigma_{xy}^i \hat{\mathcal{P}}[\dots] \sigma_{kl}^i = (n_{xl}+1) \hat{\mathcal{P}}[\dots n_{xl}+1 \dots n_{yk}-1 \dots]\Theta(n_{yk}),
\label{eq.defsinglearrowconnecting}
\end{equation}
for a connecting arrow, where $\Theta(n)$ is equal to one for $n > 0$ and zero otherwise. Here we give only the changed numbers $n_{xl}$ and $n_{yk}$. Applying the density matrix to these equations and taking the trace results again in equations of motion. Using the two types of arrows it is possible to construct every permutationally symmetric multi-level system Liouville space operator. The PsiQuaSP functions for adding one of these arrows to a given matrix are \texttt{AddMLSSingleArrowNonconnecting(...)} and \texttt{AddMLSSingleArrowConnecting(...)}. The sketch representation for the two types is shown in Fig. \ref{fig.unamsketchintro} a) and d): Equation \eqref{eq.defsinglearrownonconnecting} describing nonconnecting arrows can represent two different types of processes depending on the corresponding prefactor in the master equation. If the prefactor is imaginary the term corresponds to a Hamiltonian part $H_0$, or if it is negative and real it corresponds to dephasing, caused e.g. by a dissipator (Fig. \ref{fig.unamsketchintro} b) and c)). The two arrows are the looped and the outward pointing arrows in Fig. \ref{fig.unamsketchintro} a). 
The connecting arrow equation \eqref{eq.defsinglearrowconnecting} usually also represents two different processes: one sided flip operator actions arising from interaction Hamiltonians (Fig. \ref{fig.unamsketchintro} e) and f)) and density relaxation caused by individual spontaneous emission, decay dissipators (Fig. \ref{fig.unamsketchintro} g)).\\
\emph{Constructing physical operators:}
Looking at the collective flip operator acting from the right
\begin{align}
\mathbf{tr}[\hat{\mathcal{P}}[\dots] \rho J_{xy}]  &= \mathbf{tr}[J_{xy}\hat{\mathcal{P}}[\dots] \rho] = \mathbf{tr}[ \sum_i  \sigma_{xy}^i \hat{\mathcal{P}}[\dots] \sum_{k} \sigma_{kk}^i \rho] \nonumber\\
  &= \sum_k (n_{xk}+1) \mathcal{P}[\dots n_{xk}+1 \dots n_{yk}-1 \dots]\Theta(n_{yk})
 \label{eq.defrightflipdef}
\end{align}
amounts to summing over all possible individual connecting arrows, see Fig. \ref{fig.unamsketchintro} e). Here in the second line we have inserted the Hilbert space identity for each individual $d+1$-level system
\begin{equation}
I^i = \sum_{k} \sigma_{kk}^i.
\end{equation}
The action of the $\sigma_{xy}^i$ matrices in equation \eqref{eq.defrightflipdef} change each individual spin matrix $\sigma_{yk}^i$ into a spin matrix $\sigma_{xk}^i$. The $k$ sum of the $\sigma_{kk}^i$ matrices results in a sum over all possible right $k$ indices in $n_{yk}$ and $n_{xk}$. In the last step we insert equation \eqref{eq.defsinglearrowconnecting} and perform the trace operation. In this expression we see that the resulting matrix is sparse: The equation corresponds to the product of one row of the matrix with the column vector density matrix and thus there are at most $k$ nonzero entries in each row of this matrix.\\
The same operator acting from the left results in a sum over all possible left $k$ indices
\begin{align}
\mathbf{tr}[\hat{\mathcal{P}}[\dots] J_{xy} \rho ]   &= \mathbf{tr}[\sum_{k} \sum_i  \sigma_{kk}^i \hat{\mathcal{P}}[\dots] \sigma_{xy}^i \rho] = \sum_{k} (n_{ky}+1) \mathcal{P}[\dots n_{ky}+1 \dots n_{kx}-1 \dots]\Theta(n_{yk}).
 \label{eq.defleftflipdef}
\end{align}
These two operators can be implemented by repeatedly calling the \texttt{AddMLSSingleArrowConnecting(...)} function -- once for every possible $k$ value, see Fig. \ref{fig.unamsketchintro} e) and f). The action of a collective projection or diagonal operator $J_{xx}$ is given by 
\begin{align}
\mathbf{tr}[\hat{\mathcal{P}}[\dots] \rho J_{xx}] &= \mathbf{tr}[\sum_k \sum_i  \sigma_{xx}^i \hat{\mathcal{P}}[\dots] \sigma_{kk}^i \rho] =\sum_k n_{xk} \mathcal{P}[\dots]
 \label{eq.defrightdiagdef}
\end{align}
and
\begin{align}
\mathbf{tr}[\hat{\mathcal{P}}[\dots] J_{xx}\rho ] &= \mathbf{tr}[\sum_k \sum_i  \sigma_{kk}^i \hat{\mathcal{P}}[\dots] \sigma_{xx}^i \rho] = \sum_k n_{kx} \mathcal{P}[\dots],
 \label{eq.defleftdiagdef}
\end{align}
which can be implemented by repeatedly calling \texttt{AddMLSSingleArrowNonconnecting(...)} -- again once for every possible $k$ value, see Fig. \ref{fig.unamsketchintro} b) and c). With this set of results it is clear how to construct a general self energy Hamiltonian $\dot{\rho} \sim i/\hbar [\rho,H_0]$ or a general individual dissipator
\begin{equation}
\mathcal{D} \rho = \frac{\gamma}{2}\big( \sum_i \sigma_{xy}^i \rho \sigma_{yx}^i -J_{yy} \rho - \rho J_{yy}\big),
\end{equation}
where the first term is set by a single call to \texttt{AddMLSSingleArrowConnecting(...)}, see equation. \eqref{eq.defsinglearrowconnecting}, and the second and third term are set as in equations \ref{eq.defrightdiagdef} and \ref{eq.defleftdiagdef}. Please note that the possibility of a decoupling of some coherence degrees of freedom as in Fig. \ref{fig.mlsintro} d) is the main reason why PsiQuaSP does not provide generalized setup functions for operator actions of $J_{xy}$ and $J_{xx}$, since it would result in unnecessary numerical cost, if the decoupled basis elements were included. The other reason is that the elementary arrow representation also provides maximal freedom for the application programmer, whereas any encapsulation/facilitation would always be associated with a cut in generality.\\
The sketches for simple operators like $J_{xy}$ and $J_{xx}$ are easy to draw. 
\begin{table}
\centering
\begin{tabular}{clcl}
$ b \rho $ & \texttt{AddModeLeftB(...)} & $ \rho b  $ & \texttt{AddModeRightB(...)} \\
$ b^\dagger \rho $ & \texttt{AddModeLeftBd(...)} & $ \rho b^\dagger  $ & \texttt{AddModeRightBd(...)} \\
$ b^\dagger b \rho $ & \texttt{AddModeLeftBdB(...)} & $ \rho b^\dagger b  $ & \texttt{AddModeRightBdB(...)} \\
$ b b^\dagger \rho $ & \texttt{AddModeLeftBBd(...)} & $ \rho b b^\dagger  $ & \texttt{AddModeRightBBd(...)} \\
$ b  \rho b^\dagger $ & \texttt{AddModeLeftBRightBd(...)} & $ b^\dagger \rho b   $ & \texttt{AddModeLeftBdRightB(...)} \\
\hline
\end{tabular}
\caption{List off all available functions for setting elementary mode Liouvillians. The redundant functions allow faster and easier code development -- actually all Liouvillians could be constructed from the first two.}
\label{tab.boson}
\end{table}
\begin{figure}
\centering\includegraphics[scale=0.6]{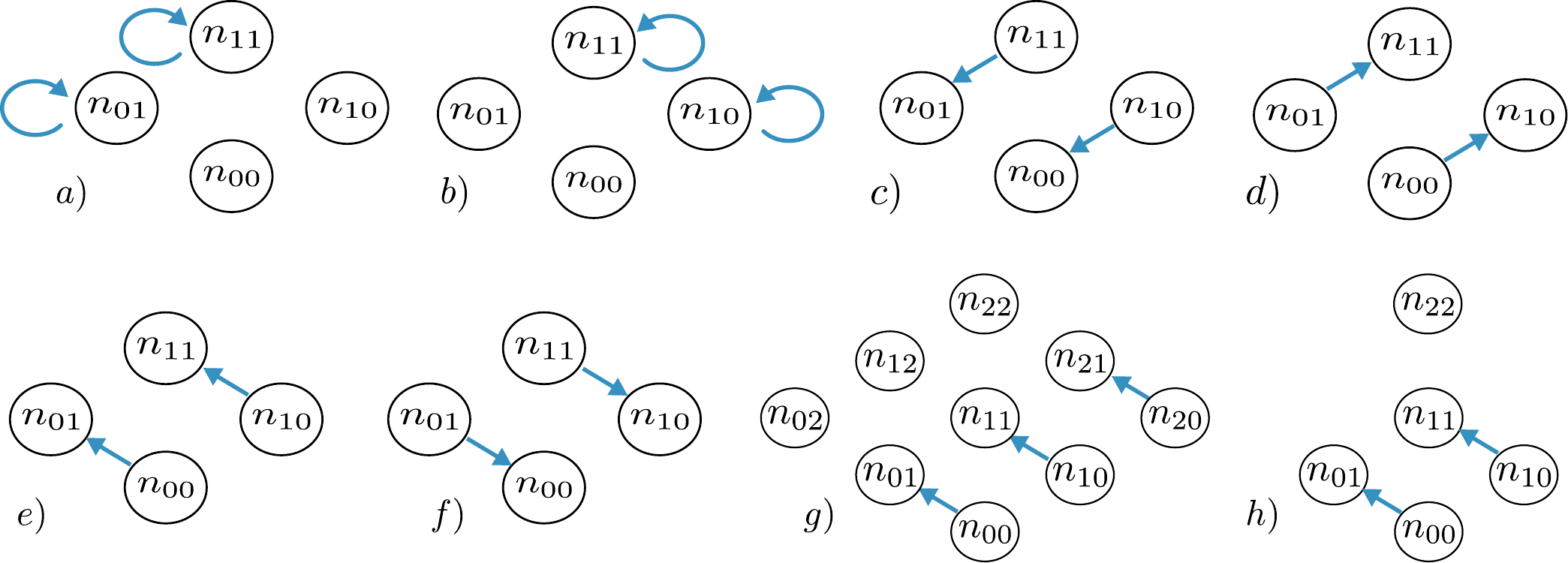}
\caption{From a) to f): Sketches corresponding to $J_{11}^{L}$, $J_{11}^{R}$, $J_{10}^{R}$, $J_{01}^{R}$ and $J_{10}^{L}$, $J_{01}^{L}$. When the operator acts on the left (right) side of the density matrix, it acts on the right (left) index of the $n_{xy}$, c.f. equation \eqref{eq.rowdef}. Two versions of the $J_{10}^L$ operators for the full and reduced three level system dynamics, c.f Fig. \ref{fig.mlsintro} c) and d).}
\label{fig.modularexample}
\end{figure}
Sketches corresponding to Liouville operators like $J_{xy}\rho J_{yx}$ or $J_{xy}^n \rho$ are more complicated and it is not recommended to implement them by hand. Rather we recommend to define the elementary processes like $J_{xy}$ and $J_{xx}$, set the corresponding matrices and then use the PETSc tools \texttt{MatMatMult()} and \texttt{MatAXPY(...)} to construct more complicated operators. The following identities are useful for this case
\begin{align}
 A \rho B ~ \widehat{=} ~ A^L\cdot B^R\rho = B^R\cdot A^L\rho,\quad\quad A B \rho ~ \widehat{=} ~ A^L\cdot B^L \rho, \quad\quad  \rho A B  ~ \widehat{=} ~ B^R\cdot A^R\rho,
\end{align}
where the $\cdot$ operation is given be the \texttt{MatMatMult()} operation.
The elementary Liouville space operators for the bosonic modes are set by calling the functions shown in Table \ref{tab.boson}.\\
\emph{Simple example:} In \texttt{example/ex4a} we implement the phonon laser/laser cooling master equation from Refs. \citenum{Kabuss:PhysRevLett:12} and \citenum{Droenner:arXiv:17}, which represents a set of two-level systems coupled to a phonon mode
\begin{equation}
H= \hbar \Delta J_{11} + \hbar \omega_{ph} b^\dagger b + \hbar g J_{11} (b+b^\dagger) + \hbar E (J_{10} + J_{01}).
\end{equation}
Here $\Delta = \omega_{11} -\omega_L$ is the detuning of the two-level systems from the driving laser. For positive detuning near the Stokes resonance this corresponds to laser cooling and for negative detuning at the anti-Stokes resonance this corresponds to phonon lasing. We include individual spontaneous emission and finite phonon lifetime
\begin{align}
\mathcal{D}_1(\rho) & = \frac{\gamma}{2} \sum_i (\sigma_{01}^i \rho \sigma_{10}^i - \sigma_{11}^i \rho - \rho \sigma_{11}^i), \qquad\mathcal{D}_2(\rho) = \frac{\kappa}{2} (b \rho b^\dagger - b^\dagger b \rho - \rho b^\dagger b).
\end{align}
In this example we need to define six two-level system operators to construct the Hamiltonian: $J_{11}^{L,R}$, $J_{10}^{L,R}$ and $J_{01}^{L,R}$. Each of these matrices is defined by two calls to \texttt{AddMLSSingleArrowNonconnecting(...)} for $J_{11}^{L,R}$ and \texttt{AddMLSSingleArrowConnecting(...)} for $J_{10}^{L,R}$ and $J_{01}^{L,R}$. The sketches for these matrices are shown in Fig. \ref{fig.modularexample}. From these matrices and the respective phonon matrices we can construct all Hamiltonians.

\section{Performance}
\label{sec.performance}

\begin{figure}
\centering
\includegraphics[scale=1]{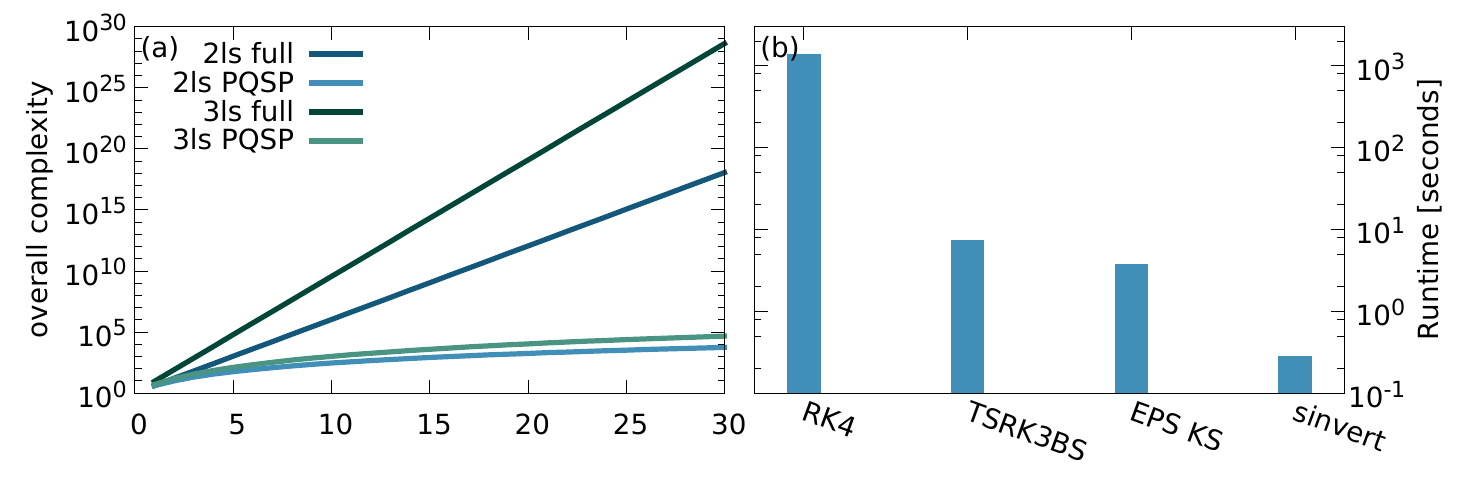}
\caption{a) The scaling of storage space and overall computation time for two-level systems and three-level systems using the full exponential approach vs. the permutation symmetric PsiQuaSP approach. b) Runtime comparison between different solution methods for steady state calculations for a two-level laser setup: fixed time step fourth order Runge-Kutta (RK4), adaptive time step Runge-Kutta (TSRK3BS), SLEPc Krylov-Schur null space computation (EPS KS) and SLEPc Krylov-Schur null space computation with exact shift and invert spectral transformation (sinvert). Please refer to the PETSc and SLEPc documentation for details to these solvers.}
\label{fig.perf}
\end{figure}

The two main advantages of PsiQuaSP are the reduction of complexity due to the symmetrized basis states and the manifold of solvers provided through PETSc and e.g. SLEPc.\\
\emph{Overall complexity:} In Fig. \ref{fig.perf} a) the number of basis elements of the density matrix for the full exponential density matrix is compared to the polynomial, symmetrized PsiQuaSP density matrix for two- and three-level systems. This corresponds to the overall complexity since both the storage requirement and the number of coupled equations scale like the number of basis elements.\\
\emph{Steady state computation:} In Fig. \ref{fig.perf} b) the convergence time for steady state calculations for a two-level laser as discussed in Ref. \citenum{Gegg:NJP:16} and implemented in the examples \texttt{example/ex2a} and \texttt{example/ex2b} for different solvers is shown: the fixed time step fourth order Runge-Kutta is by far the slowest solver. The adaptive time step and the direct null space computation using the SLEPc package outperform this standard routine. The speedup of the shift and invert spectral transformation solver \cite{petsc-user-ref,petsc-efficient,Hernandez:TMS:05,slepc-users-manual} compared to the RK4 method amounts almost to a factor of $5000$. Please note that these numbers and the relative performance of these solvers is parameter and system size dependent, it is possible to find examples where the difference is even higher but it is also possible to find examples where the difference is less pronounced. Especially for iterative solvers like the SLEPc Krylov-Schur eigenvalue solver convergence time is highly dependent on the spectrum of the matrix and on chosen solver specific parameters. Please refer to the PETSc and SLEPc documentations for the specifics of these methods.

\section{Summary}
We have introduced a library that enables the setup of master equations for identical multi-level systems. The library provides ready-made setup functions for density matrices as well as Liouville operators. The design of these functions is centered around the sketch representation of the Liouville operators or master equation introduced in Ref. \citenum{Gegg:NJP:16}. This has the advantage that implementing an arbitrary master equation does not require calculating any equations of motion but can be done by directly implementing the sketches. There is a simplified usage for two-level systems and ready-made Liouvillian steup routines and an advanced usage where the user can construct arbitrary permutation symmetric Liouvillians from simple sketches.

\section{Acknowledgements}
We gratefully acknowledge funding of the Deutsche Forschungsgemeinschaft (DFG) through SFB 951 (M.G, M.R) and through the School of Nanophotonics of SFB 787 (M.G.). We further want to thank Andreas Knorr for useful discussions and thank Christopher W\"achtler and Leon Droenner for helping with benchmarking and bug fixing.

\section{Author contributions}
M. G. wrote the manuscript and the code, both authors conceived the methodology and edited the manuscript.

\section{Additional information}
\subsection*{PsiQuaSP code:}
The code of the PsiQuaSP library can be found on GitHub: \href{https://github.com/modmido/psiquasp}{https://github.com/modmido/psiquasp}

\subsection*{Competing interests:}
The authors declare no competing financial interests.

\end{document}